\documentclass{jetpl_arXiv13021496}
\twocolumn

\newcommand{\beq}{\begin{equation}}
\newcommand{\eeq}{\end{equation}}
\newcommand{\beqa}{\begin{eqnarray}}
\newcommand{\eeqa}{\end{eqnarray}}
\newcommand{\bsubeqs}{\begin{subequations}}
\newcommand{\esubeqs}{\end{subequations}}

\title{\vspace*{4mm}
Standard Model Higgs field and energy scale of gravity} 

\rtitle{Standard Model Higgs field and energy scale of gravity} 

\sodtitle{}

\author{F.R. Klinkhamer $^{\#}$
\/\thanks{\:e-mail: frans.klinkhamer@kit.edu}}

\rauthor{F.R. Klinkhamer}

\sodauthor{Klinkhamer}

\address{$^{\#}$ Institute for Theoretical Physics,
Karlsruhe Institute of Technology (KIT), 76128 Karlsruhe, Germany}


\abstract{The effective potential of the Higgs scalar field in the Standard Model
may have a second degenerate minimum at an ultrahigh vacuum expectation value.
This second minimum then determines, by radiative corrections,
the values of the top-quark and Higgs-boson masses
at the standard minimum corresponding to the electroweak energy scale.
An argument is presented that this ultrahigh vacuum expectation value
is proportional to the energy scale of gravity,
$E_\text{Planck} \equiv \sqrt{\hbar\, c^5/G_{N}}$,
considered to be characteristic of a spacetime foam.
In the context of a simple model, the existence of
kink-type wormhole solutions places a lower bound on the
ultrahigh vacuum expectation value and this lower bound
is of the order of $E_\text{Planck}$.
\vspace*{-3mm}}

\begin{document}

\maketitle

\section{INTRODUCTION}\label{sec:Introduction}

The ATLAS and CMS collaborations
have recently reported results~\cite{Higgs-ATLAS2012,Higgs-CMS2012}
which appear to confirm the existence
of the Higgs boson of the Standard Model with a mass around
$126\;\text{GeV}/c^2$.
Long before that, Froggatt and Nielsen~\cite{FroggattNielsen1995}
gave a remarkable prediction of the Higgs mass value
($M_\text{Higgs}= 135 \pm 9 \;\text{GeV}/c^2$) based on a heuristic
physical argument, multiple-point criticality.
A crucial ingredient of the prediction was the following identification
for the ultrahigh vacuum expectation value
at a second degenerate minimum of the effective Higgs potential:
\vspace*{-1mm}
\beq\label{eq:phi2}
\phi_\text{vac,2}
\stackrel{?}{\sim}
E_\text{Planck}\equiv \sqrt{\hbar\, c^5/G_{N}}
\approx 1.22 \times 10^{19}\;\text{GeV}\,,       
\eeq
with the first minimum corresponding to the standard electroweak
scale, $\phi_\text{vac,1} \approx 246\;\text{GeV}$.

But the physics motivation for the identification \eqref{eq:phi2}
was indirect and \eqref{eq:phi2} was really only an assumption.
In fact, Froggatt and Nielsen did not calculate
gravitational effects
(governed by Newton's coupling constant $G_N$)
but simply appealed to the relevance of Planckian units
as a \textit{deus ex machina}.
Obviously, it would be conceptually important to understand why
$\phi_\text{vac,2} \propto 1/\sqrt{G_N}$ and to see that the
proportionality constant in \eqref{eq:phi2} is indeed of order $1$.
Related issues have also been addressed in several recent
papers (see, e.g.,
Refs.~\cite{MeissnerNicolai2006,ShaposhnikovWetterich2009,Holthausen-etal2011}),
but our approach is different.

It is expected (but, of course, not proven) that the fundamental
structure of spacetime changes radically at energies of order
$E_\text{Planck}$ or length scales of the order of
$\hbar c/E_\text{Planck}$.
Over the years, various aspects of this radical change   
have been considered, going under the names
of spacetime foam, superstrings, and loop quantum gravity.
As the aim of the present paper is rather modest,
we will restrict ourselves to a very simple model and a very
simple calculation.
Hopefully, this will give us at least one physical argument
of what may determine the parametric dependence of $\phi_\text{vac,2}$.

\section{SETUP}\label{sec:Setup}

In place of the full Standard Model at typical interaction energies and renormalization scale
of order $E_\text{Planck}$
(described in part by the effective
potential~\cite{ColemanWeinberg1973,Sher1988,Ford-etal1992,ChetyrkinZoller2012,%
Bezrukov-etal2012,Degrassi-etal2012}),
consider  a simple classical theory with
a single scalar field and Einstein gravity.
[Time scales of the  
classical theory, or length scales divided by $c$, are converted  
into inverse-energy scales by the
introduction of the Planck constant $\hbar$\,.]
Concretely, take  
\vspace*{0.35\baselineskip}\newline  
\hspace*{5mm}$\bullet$
a real classical scalar field $\phi(x)$;
\vspace*{0.2\baselineskip}\newline
\hspace*{5mm}$\bullet$
a scalar potential $V(\phi)\geq 0$ with two degenerate\newline
\hspace*{5mm}$\phantom{\bullet}$
minima, $V(v_1)=V(v_2)=0$;
\vspace*{0.2\baselineskip}\newline
\hspace*{5mm}$\bullet$
a conformal coupling of the scalar field to gravity\newline
\hspace*{5mm}$\phantom{\bullet}$
(coupling constant $\xi=1/6$).
\vspace*{-0.15\baselineskip}\newline
\par
Our goal, now, is to perform a toy-model calculation of something like a
spacetime foam. The easiest calculation is to look for
permanent static Lorentzian wormholes~\cite{Visser1995}.
For the simple classical theory considered,
Sushkov and Kim~\cite{SushkovKim2002}
have indeed found regular kink-type wormhole solutions. Remarkably,
these solutions only occur for the case of `small' $v_1$
and `large' $v_2\,$:
\beq\label{eq:wormhole-condition}
|v_{1}| < E_\text{Planck}/\sqrt{8\pi\xi}\,,\quad
|v_{2}| > E_\text{Planck}/\sqrt{8\pi\xi}\,,
\eeq
which can be written more compactly in terms of
the so-called reduced Planck energy,
$E_\text{P} \equiv E_\text{Planck}/ \sqrt{8\pi}$.
The heuristic explanation of \eqref{eq:wormhole-condition}
is that for this case the conformal factor
$f(\phi) \equiv 1-8\pi \xi\, \phi^2/(E_\text{Planck})^2$
of a kink-type scalar field configuration $\phi(\rho)$
can vanish for one and only one value
$\rho_{0}$ of the radial coordinate $\rho$,
whereas pairs of such points, $\rho_{0,1}$ and $\rho_{0,2}$, would have
a nondifferentiable solution in between  
[this conformal factor $f(\phi)$ multiplies the Ricci scalar
$R$ in the action and further details can be found below].

Note the crucial role of having finite positive $\xi$
in \eqref{eq:wormhole-condition} and the
possibly convincing argument in favor of the
value $\xi=1/6$ from conformal symmetry (see, e.g., the discussion
in Ref.~\cite{MeissnerNicolai2006}).

But before investigating the implications  
of \eqref{eq:wormhole-condition}
for the electroweak theory, we must make sure that a wormhole  
solution still exists if $|v_1| \ll |v_2|$.

\section{MODEL}\label{sec:Model}  

We consider the following classical model
(setting $G_{N}=c=\hbar=1$ and using the same conventions as in
Ref.~\cite{SushkovKim2002}):
\bsubeqs\label{eq:model}
\beqa\label{eq:model-action}
S &=&
\int{}d^4x\,\sqrt{-g}\;
\Big[ \frac{1}{16\pi}\,R
-\frac{1}{2}\, g^{\mu\nu}\, \phi_{;\mu} \phi_{;\nu}
\nonumber\\
&&
-\frac{1}{2}\, \xi\, \phi^2\, R - V(\phi)\Big]\,,
\\[2mm]\label{eq:model-pot}
V(\phi) &=& \frac{\lambda}{4}\, \big(\phi -v_1\big)^2 \,\big(\phi -v_2\big)^2 \,,
\\[2mm]\label{eq:model-v1-v2-order}
\lambda &>& 0\,,\quad  0 \leq   v_1 <  v_2 \,,
\\[2mm]\label{eq:model-xi}
\xi &=& 1/6\,.
\eeqa
\esubeqs
A more realistic potential would involve logarithms of $\phi^2$
(cf. Refs.~\cite{ColemanWeinberg1973,Sher1988,Ford-etal1992}), but the
polynomial potential \eqref{eq:model-pot} is used for simplicity.

Following Ref.~\cite{SushkovKim2002}, the
spherically symmetric static \textit{Ansatz} is given by
\bsubeqs\label{eq:Ansatz}
\beqa
ds^2 &=& - A(\rho)\, dt^2 + \frac{d\rho^2}{A(\rho)}
+ \widehat{r}^{\,2}(\rho)\,\big(d\theta^2
+ \sin^2 \theta \,  d\phi^2\big)\,,\\[2mm]
\phi &=& \phi(\rho)\,.
\eeqa
\esubeqs
At this moment, it turns out to be useful to introduce further
model parameters:
\bsubeqs\label{eq:parameters}
\beqa
\kappa &\equiv& m/\sqrt{\lambda} \equiv (v_2-v_1)/2 \,,\\[2mm]
\overline{\phi} &=& (v_2+v_1)/2\,,
\eeqa
\esubeqs
and the following dimensionless variables:
\bsubeqs\label{eq:variables} 
\beqa
y               &\equiv&  \frac{m\, \rho}{\sigma +|m\, \rho|} \,, \\[2mm]
r(y)            &\equiv& m \,\widehat{r}(\rho) \,, \\[2mm]
\eta(y)         &\equiv&  \phi(\rho)/\kappa \,, \\[2mm]
\overline{\eta} &\equiv&  \overline{\phi}/\kappa \,,
\eeqa
\esubeqs
with a positive numerical constant $\sigma$ in the definition
of the compactified dimensionless radial coordinate $y$.
The minima of the potential \eqref{eq:model-pot} then occur
for the following vacuum expectation values
of the dimensionless scalar field:
\beq
\eta_1 = \overline{\eta}-1\,,\quad
\eta_2 = \overline{\eta}+1.
\eeq

For the above \textit{Ansatz} and definitions,
the reduced field equations and boundary conditions
are given by Eqs.~(4.34)--(4.39) in Ref.~\cite{SushkovKim2002},
where a typo in the definition of $f$ stands to be corrected.
These reduced field equations can only be solved numerically.

\section{NUMERICAL SOLUTION}\label{sec:Numerical-solution}

The authors of
Ref.~\cite{SushkovKim2002} have presented a numerical solution
(also reproduced by us) for a particular set of parameters and
boundary conditions, having, in particular,
scalar minima $\eta_1 \approx 1.4495 $ and $\eta_2 \approx 3.4495$
for model parameter $\overline{\eta} = \sqrt{6}$.
The corresponding dimensional vacuum expectation values
$v_1$ and $v_2$ are both Planckian, whereas we are interested
in having one, $v_1$, at the electroweak scale.

\begin{figure*}[t]  
\vspace*{0mm}
\begin{center}             
\includegraphics[width=0.75\textwidth]{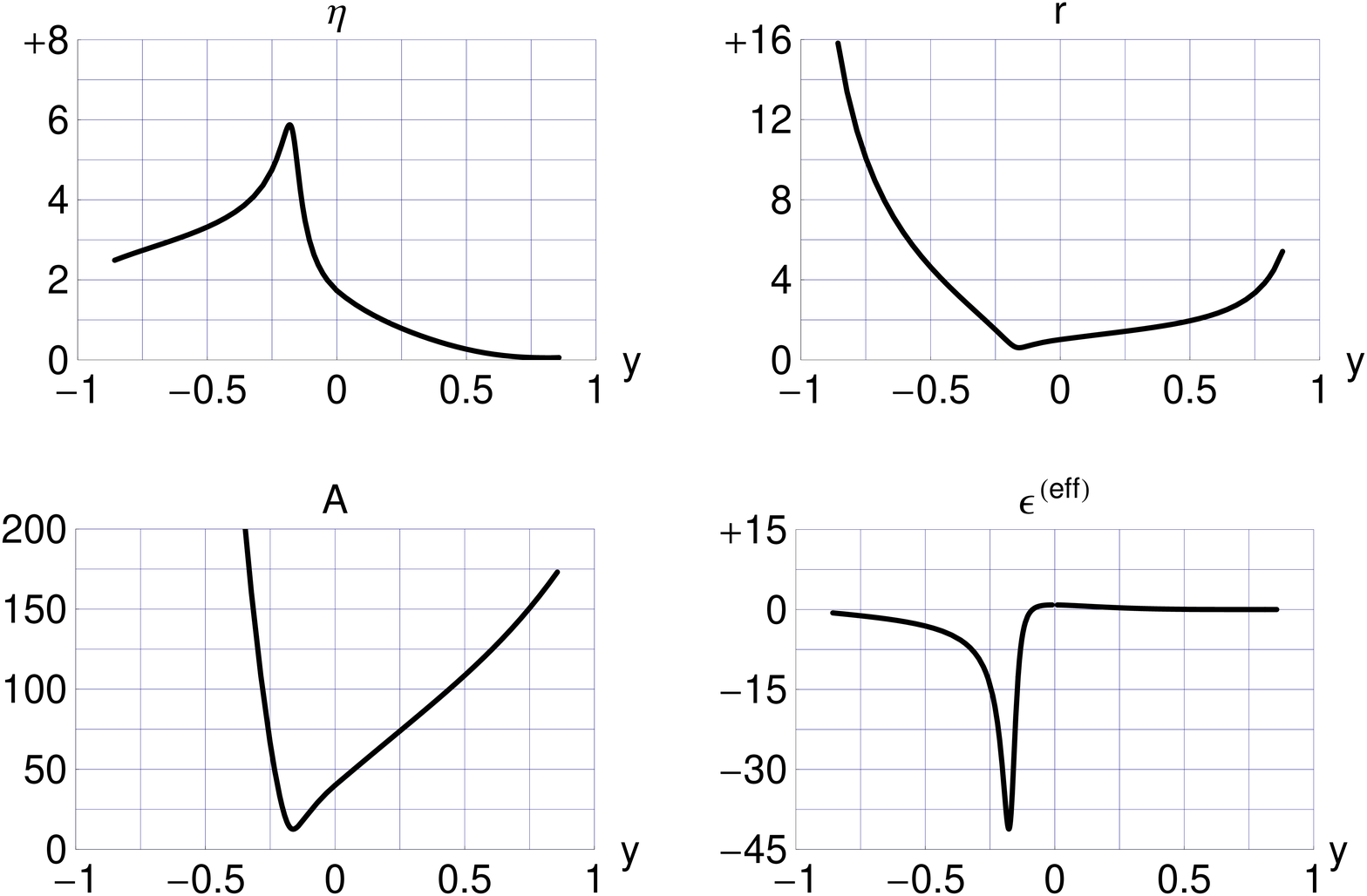}
\end{center}
\vspace*{-1mm}
\caption{Fig.~1:\, Numerical solution of the reduced field equations from
the spherically symmetric \textit{Ansatz} \eqref{eq:Ansatz}
in the scalar-gravity model \eqref{eq:model}. The top-left panel shows      
the dimensionless scalar field $\eta$ as a function of the
compactified dimensionless radial coordinate $y$. The scalar field
$\eta(y)$ is seen to interpolate between the vacuum values
$\eta_1 = 0$ and $\eta_2 =2$, making for a non-monotonic
kink-type configuration. The further panels
show the dimensionless metric functions $A(y)$ and $r(y)$
and the dimensionless effective energy density
$\epsilon^\text{(eff)}[\eta(y),\,  A(y),\, r(y)]$.
The dimensionless `radius' function $r(y)$ is seen to have a minimum
value of approximately $0.6$ at radial coordinate
$y \approx -0.16$, which corresponds to the so-called wormhole
throat~\cite{Visser1995}.
The model parameters used are
$\{\xi,\, \lambda,\, \kappa,\,\overline{\eta},\,\sigma\}$
$=$ $\{1/6,\,1,\,  1/(2\sqrt{\pi}),\,  1,\, 10\}$.
For the numerical solution of the reduced field equations
[three second-order ordinary differential equations],
the boundary conditions at $y=0$ are
$\{r(0),\, A(0),\, \eta(0),\, \eta^\prime(0)\}$ $=$
$\{1017/1000,\,  40,\,  \sqrt{3},\,  -3/5\}$,
with $r^\prime(0)$ and $A^\prime(0)$ values from two constraint
equations (see Ref.~\cite{SushkovKim2002} for details).\vspace*{-0mm}}
\label{fig:SM-type-wormhole}
\end{figure*}

We have, therefore, obtained a
numerical solution for $\eta_1 = 0$ and $\eta_2 = 2$;
see the caption of Fig.~\ref{fig:SM-type-wormhole} 
for the specific parameters and boundary conditions used
(the conformal factor $1-8\pi\xi\kappa^2\eta^2$ vanishes at $y=0$).
The resulting spacetime and scalar field configuration (Fig.~1)  
can be described as follows:
\begin{itemize}
  \item
on the `outside' of the wormhole
($y > y_\text{throat} \approx -0.16$),  
there is a smooth approach to the  standard Minkowski spacetime and
the Standard Model Higgs vacuum $\phi =v_1$.
\item
on the `inside' of the wormhole ($y < y_\text{throat}$),
there is a Planck-scale scalar field
$\phi \sim v_2$ with effective
energy densities of order $-(E_\text{Planck})^4$ close
to the wormhole throat, which may be viewed as a caricature of
what a dynamical quantum spacetime foam can look like
at ultrashort length scales.
\end{itemize}
The results shown in Fig.~\ref{fig:SM-type-wormhole} 
can be expected to give
a reasonably accurate approximation of the
exact wormhole-type solution over the coordinate interval
$-0.5  \lesssim y \lesssim 0.75$.

\section{DISCUSSION}\label{sec:Discussion}

In view of these numerical results and in
line with condition \eqref{eq:wormhole-condition},
the first wormhole solutions
of model \eqref{eq:model} 
would occur for
\vspace*{-0mm}
\bsubeqs\label{eq:v1-v2-values}
\beqa\label{eq:v1-value}
v_1 &\ll& E_\text{Planck}\,,
\\[2mm]\label{eq:v2-value}
v_2 &\sim&
\sqrt{\frac{1}{8\pi \xi}}\;E_\text{Planck}
=
\frac{1}{2}\,\sqrt{\frac{3}{\pi}}\;E_\text{Planck}
\nonumber\\
&\approx&
5.97 \times 10^{18}\;\text{GeV}\,, 
\eeqa
\esubeqs
where the conformal value \eqref{eq:model-xi} for the
coupling constant $\xi$ has been used in the last step.
Now, this is indeed what may be relevant for
the renormalization-group-improved effective potential of the
Standard Model~\cite{ColemanWeinberg1973,Sher1988,Ford-etal1992}
entering the multiple-point-criticality argument
of Froggatt and Nielsen~\cite{FroggattNielsen1995},
with $v_1 \sim  10^{2}\;\text{GeV}$ and
$v_2 \sim 10^{19}\;\text{GeV}$.

Taking \eqref{eq:v2-value}
at face value and extrapolating one set of NNLO results from the
right panel of Fig.~4 in Ref.~\cite{Degrassi-etal2012}
gives the following pole masses:  
$M_\text{Higgs} = 126\;\text{GeV}$ and
$M_\text{top} \approx 171.4\;\text{GeV}$,
for $\alpha_{s}(M_Z)=0.1184$. With input values
$M_\text{Higgs} \in [124\;\text{GeV},\, 128\;\text{GeV}]$
and $\alpha_{s}(M_Z)\in [0.1160,\, 0.1210]$, there is
the following prediction by linear approximation:
$M_\text{top}[\text{GeV}] \approx 171.4
+\big(M_\text{Higgs}[\text{GeV}] - 126\big)\big/2
+\big(\alpha_{s}(M_Z)-0.1184\big)\big/0.0028$,
with an estimated theoretical $1\sigma$ uncertainty of $\pm 0.5$
(see Ref.~\cite{Degrassi-etal2012}  for details
and further discussion of technical issues).

We repeat that the
simple classical model \eqref{eq:model} is only considered to
describe certain aspects of the Standard Model physics
at typical interaction energies and renormalization scale
of order $E_\text{Planck}$ (observe, for example, that the curvature
around the $v_1$ minimum has a Planckian order of magnitude,
contrary to what is observed
experimentally~\cite{Higgs-ATLAS2012,Higgs-CMS2012}).
Still, the Standard Model fields may suffice to explain all particle physics
results known to date, including neutrino masses and mixing
(the dimension-5 term discussed in Ref.~\cite{Klinkhamer2011}
would have a mass scale $M_5 \sim v_2/c^2$).

Let us close with two remarks.
First, there is, in principle, no problem to extend the
\textit{Ansatz} \eqref{eq:Ansatz} of the simple model
to the Standard Model fields, having made an obvious generalization
of the degenerate potential \eqref{eq:model-pot}
and adding appropriate spherically symmetric
gauge fields.  Assuming that a regular solution exists,
the next issue is stability.
We are moderately optimistic because the existence and stability
of the flat-spacetime kink solution in $1+1$ dimensions
does not require gauge fields in the
first place (different from the Nielsen--Olesen vortex solution
in the Abelian $U(1)$ Higgs model, which, in fact,
looses its stability when embedded in the Standard Model~\cite{KlinkhamerOlesen1994}).

Second, indirect (Cherenkov) experimental bounds~\cite{BernadotteKlinkhamer2006}
require a sufficiently dilute gas of static wormholes
as considered in this paper. But, if there exist
indeed wormhole-type spacetime defects, they are,
most likely, nonstatic and without preferred frame.
The simple type of wormhole solution considered here is only
for the purpose of determining the parametric behavior of
the ultrahigh vacuum expectation value of a second degenerate
minimum of the effective Higgs potential.

\vspace*{-6mm}
\section*{ACKNOWLEDGMENTS}\noindent
The author is grateful to M. Schwarz for pointing out Ref.~\cite{SushkovKim2002},
already several years ago.

\end{document}